
----------
X-Sun-Data-Type: default
X-Sun-Data-Name: oddmailer.tex
X-Sun-Content-Lines: 1386

\message
{JNL.TEX version 0.9 as of 3/27/86.  Report bugs and problems to Doug Eardley.}



\font\twelverm=cmr10 scaled 1200    \font\twelvei=cmmi10 scaled 1200
\font\twelvesy=cmsy10 scaled 1200   \font\twelveex=cmex10 scaled 1200
\font\twelvebf=cmbx10 scaled 1200   \font\twelvesl=cmsl10 scaled 1200
\font\twelvett=cmtt10 scaled 1200   \font\twelveit=cmti10 scaled 1200

\skewchar\twelvei='177   \skewchar\twelvesy='60


\def\twelvepoint{\normalbaselineskip=12.4pt plus 0.1pt minus 0.1pt
  \abovedisplayskip 12.4pt plus 3pt minus 9pt
  \belowdisplayskip 12.4pt plus 3pt minus 9pt
  \abovedisplayshortskip 0pt plus 3pt
  \belowdisplayshortskip 7.2pt plus 3pt minus 4pt
  \smallskipamount=3.6pt plus1.2pt minus1.2pt
  \medskipamount=7.2pt plus2.4pt minus2.4pt
  \bigskipamount=14.4pt plus4.8pt minus4.8pt
  \def\rm{\fam0\twelverm}          \def\it{\fam\itfam\twelveit}%
  \def\sl{\fam\slfam\twelvesl}     \def\bf{\fam\bffam\twelvebf}%
  \def\mit{\fam 1}                 \def\cal{\fam 2}%
  \def\tt{\twelvett}
  \textfont0=\twelverm   \scriptfont0=\tenrm   \scriptscriptfont0=\sevenrm
  \textfont1=\twelvei    \scriptfont1=\teni    \scriptscriptfont1=\seveni
  \textfont2=\twelvesy   \scriptfont2=\tensy   \scriptscriptfont2=\sevensy
  \textfont3=\twelveex   \scriptfont3=\twelveex  \scriptscriptfont3=\twelveex
  \textfont\itfam=\twelveit
  \textfont\slfam=\twelvesl
  \textfont\bffam=\twelvebf \scriptfont\bffam=\tenbf
  \scriptscriptfont\bffam=\sevenbf
  \normalbaselines\rm}



\def\beginlinemode{\endmode
  \begingroup\parskip=0pt \obeylines\def\\{\par}\def\endmode{\par\endgroup}}
\def\beginparmode{\endmode
  \begingroup \def\endmode{\par\endgroup}}
\let\endmode=\par
{\obeylines\gdef\
{}}
\def\singlespace{\baselineskip=\normalbaselineskip}

\def\oneandahalfspace{\baselineskip=\normalbaselineskip
  \multiply\baselineskip by 3 \divide\baselineskip by 2}
\def\doublespace{\baselineskip=\normalbaselineskip \multiply\baselineskip by 2}

\newcount\firstpageno
\firstpageno=2
\footline={\ifnum\pageno<\firstpageno{\hfil}\else{\hfil\twelverm\folio\hfil}\fi}
\def\toppageno{\global\footline={\hfil}\global\headline
  ={\ifnum\pageno<\firstpageno{\hfil}\else{\hfil\twelverm\folio\hfil}\fi}}
\let\rawfootnote=\footnote		
\def\footnote#1#2{{\rm\singlespace\parindent=0pt\parskip=0pt
  \rawfootnote{#1}{#2\hfill\vrule height 0pt depth 6pt width 0pt}}}
\def\raggedcenter{\leftskip=4em plus 12em \rightskip=\leftskip
  \parindent=0pt \parfillskip=0pt \spaceskip=.3333em \xspaceskip=.5em
  \pretolerance=9999 \tolerance=9999
  \hyphenpenalty=9999 \exhyphenpenalty=9999 }
\def\dateline{\rightline{\ifcase\month\or
  January\or February\or March\or April\or May\or June\or
  July\or August\or September\or October\or November\or December\fi
  \space\number\year}}
\def\received{\vskip 3pt plus 0.2fill
 \centerline{\sl (Received\space\ifcase\month\or
  January\or February\or March\or April\or May\or June\or
  July\or August\or September\or October\or November\or December\fi
  \qquad, \number\year)}}


\hsize=6.5truein
\vsize=8.9truein
\parskip=\medskipamount
\def\\{\cr}
\twelvepoint		
\doublespace		
\overfullrule=0pt	




\def\title			
  {\null\vskip 3pt plus 0.2fill
   \beginlinemode \doublespace \raggedcenter \bf}

\def\author			
  {\vskip 3pt plus 0.2fill \beginlinemode
   \singlespace \raggedcenter}

\def\affil			
  {\vskip 3pt plus 0.1fill \beginlinemode
   \oneandahalfspace \raggedcenter \sl}

\def\abstract			
  {\vskip 3pt plus 0.3fill \beginparmode
   \oneandahalfspace ABSTRACT: }

\def\endtopmatter		
  {\endpage			
   \body}

\def\body			
  {\beginparmode}		

\def\head#1{			
  \goodbreak\vskip 0.5truein	
  {\immediate\write16{#1}
   \raggedcenter \uppercase{#1}\par}
   \nobreak\vskip 0.25truein\nobreak}

\def\beneathrel#1\under#2{\mathrel{\mathop{#2}\limits_{#1}}}

\def\refto#1{$^{#1}$}		

\def\references			
  {\head{References}		
   \beginparmode
   \frenchspacing \parindent=0pt \leftskip=1truecm
   \parskip=8pt plus 3pt \everypar{\hangindent=\parindent}}

\gdef\refis#1{\item{#1.\ }}			

\gdef\journal#1, #2, #3, 1#4#5#6{		
    {\sl #1~}{\bf #2}, #3 (1#4#5#6)}		

\def\endreferences{\body}

\def\figurecaptions		
  {\endpage
   \beginparmode
   \head{Figure Captions}
}

\def\endpage			
  {\vfill\eject}

\def\endpaper			
  {\endmode\vfill\supereject}


\def\heading				
  {\vskip 0.5truein plus 0.1truein	
   \beginparmode \def\\{\par} \parskip=0pt \singlespace \raggedcenter}

\def\subheading				
  {\vskip 0.25truein plus 0.1truein	
   \beginlinemode \singlespace \parskip=0pt \def\\{\par}\raggedcenter}

\def\tag#1$${\eqno(#1)$$}

\def\align#1$${\eqalign{#1}$$}

\def\aligntag#1$${\gdef\tag##1\\{&(##1)\cr}\eqalignno{#1\\}$$
  \gdef\tag##1$${\eqno(##1)$$}}

\def\overset#1\to#2{{\mathop{#2}^{#1}}}
\def\underset#1\to#2{{\mathop{#2}_{#1}}}


\def\ref#1{Ref.~#1}			
\def\Ref#1{Ref.~#1}			
\def\[#1]{[\cite{#1}]}
\def\cite#1{{#1}}
\def\(#1){(\call{#1})}
\def\call#1{{#1}}
\def\taghead#1{}
\def\frac#1#2{{#1 \over #2}}

\def\12{{1\over2}}

\def\sla{\raise.15ex\hbox{$/$}\kern-.57em}
\def\leaderfill{\leaders\hbox to 1em{\hss.\hss}\hfill}
\def\twiddle{\lower.9ex\rlap{$\kern-.1em\scriptstyle\sim$}}
\def\bigtwiddle{\lower1.ex\rlap{$\sim$}}
\def\gtwid{\mathrel{\raise.3ex\hbox{$>$\kern-.75em\lower1ex\hbox{$\sim$}}}}
\def\ltwid{\mathrel{\raise.3ex\hbox{$<$\kern-.75em\lower1ex\hbox{$\sim$}}}}
\def\square{\kern1pt\vbox{\hrule height 1.2pt\hbox{\vrule width 1.2pt\hskip 3pt
   \vbox{\vskip 6pt}\hskip 3pt\vrule width 0.6pt}\hrule height 0.6pt}\kern1pt}
\def\tdot#1{\mathord{\mathop{#1}\limits^{\kern2pt\ldots}}}

\def\pmb#1{\setbox0=\hbox{#1}%
  \kern-.025em\copy0\kern-\wd0
  \kern  .05em\copy0\kern-\wd0
  \kern-.025em\raise.0433em\box0 }

\def\3he{{$^3${\rm He}}}

\def\slD{\raise.15ex\hbox{$/$}\kern-.57em\hbox{$D$}}
\def\dsl{\raise.15ex\hbox{$/$}\kern-.57em\hbox{$\Delta$}}
\def\slp{{\raise.15ex\hbox{$/$}\kern-.57em\hbox{$\partial$}}}
\def\nsl{\raise.15ex\hbox{$/$}\kern-.57em\hbox{$\nabla$}}
\def\sla{\raise.15ex\hbox{$/$}\kern-.57em\hbox{$\rightarrow$}}
\def\slla{\raise.15ex\hbox{$/$}\kern-.57em\hbox{$\lambda$}}
\def\slb{\raise.15ex\hbox{$/$}\kern-.57em\hbox{$b$}}
\def\lnp{\raise.15ex\hbox{$/$}\kern-.57em\hbox{$p$}}
\def\lnk{\raise.15ex\hbox{$/$}\kern-.57em\hbox{$k$}}
\def\lnK{\raise.15ex\hbox{$/$}\kern-.57em\hbox{$K$}}
\def\lnq{\raise.15ex\hbox{$/$}\kern-.57em\hbox{$q$}}
\def\lnA{\raise.15ex\hbox{$/$}\kern-.57em\hbox{$A$}}
\def\lna{\raise.15ex\hbox{$/$}\kern-.57em\hbox{$a$}}
\def\lnB{\raise.15ex\hbox{$/$}\kern-.57em\hbox{$B$}}

\def\a{\alpha}


\def\pmb#1{\setbox0=\hbox{$#1$}%
\kern-.025em\copy0\kern-\wd0
\kern.05em\copy0\kern-\wd0
\kern-.025em\raise.0433em\box0 }

\def\q2{{Q^2}}
\def\gtwid{\raise.3ex\hbox{$>$\kern-.75em\lower1ex\hbox{$\sim$}}}
\def\ltwid{\raise.3ex\hbox{$<$\kern-.75em\lower1ex\hbox{$\sim$}}}
\def\12{{1\over2}}
\def\part{\partial}

\def\low#1{\lower.5ex\hbox{${}_#1$}}

\def\psl{\raise.15ex\hbox{$/$}\kern-.57em\hbox{$\partial$}}
\def\partt{\raise.15ex\hbox{$\widetilde$}{\kern-.37em\hbox{$\partial$}}}

\def\topppageno1{\global\footline={\hfil}\global\headline
={\ifnum\pageno<\firstpageno{\hfil}\else{\hss\twelverm --\ \folio
\ --\hss}\fi}}

\def\toppageno2{\global\footline={\hfil}\global\headline
={\ifnum\pageno<\firstpageno{\hfil}\else{\rightline{\hfill\hfill
\twelverm \ \folio
\ \hss}}\fi}}

\catcode`@=11
\newcount\r@fcount \r@fcount=0
\newcount\r@fcurr
\immediate\newwrite\reffile
\newif\ifr@ffile\r@ffilefalse
\def\w@rnwrite#1{\ifr@ffile\immediate\write\reffile{#1}\fi\message{#1}}

\def\writer@f#1>>{}
\def\referencefile{
  \r@ffiletrue\immediate\openout\reffile=\jobname.ref%
  \def\writer@f##1>>{\ifr@ffile\immediate\write\reffile%
    {\noexpand\refis{##1} = \csname r@fnum##1\endcsname = %
     \expandafter\expandafter\expandafter\strip@t\expandafter%
     \meaning\csname r@ftext\csname r@fnum##1\endcsname\endcsname}\fi}%
  \def\strip@t##1>>{}}

\def\citeall#1{\xdef#1##1{#1{\noexpand\cite{##1}}}}
\def\cite#1{\each@rg\citer@nge{#1}}     

\def\each@rg#1#2{{\let\thecsname=#1\expandafter\first@rg#2,\end,}}
\def\first@rg#1,{\thecsname{#1}\apply@rg}       
\def\apply@rg#1,{\ifx\end#1\let\next=\relax
\else,\thecsname{#1}\let\next=\apply@rg\fi\next}

\def\citer@nge#1{\citedor@nge#1-\end-}  
\def\citer@ngeat#1\end-{#1}
\def\citedor@nge#1-#2-{\ifx\end#2\r@featspace#1 
  \else\citel@@p{#1}{#2}\citer@ngeat\fi}        
\def\citel@@p#1#2{\ifnum#1>#2{\errmessage{Reference range #1-#2\space is bad.}%
    \errhelp{If you cite a series of references by the notation M-N, then M and
    N must be integers, and N must be greater than or equal to M.}}\else%
 {\count0=#1\count1=#2\advance\count1
by1\relax\expandafter\r@fcite\the\count0,%

  \loop\advance\count0 by1\relax
    \ifnum\count0<\count1,\expandafter\r@fcite\the\count0,%
  \repeat}\fi}

\def\r@featspace#1#2 {\r@fcite#1#2,}    
\def\r@fcite#1,{\ifuncit@d{#1}
    \newr@f{#1}%
    \expandafter\gdef\csname r@ftext\number\r@fcount\endcsname%
                     {\message{Reference #1 to be supplied.}%
                      \writer@f#1>>#1 to be supplied.\par}%
 \fi%
 \csname r@fnum#1\endcsname}
\def\ifuncit@d#1{\expandafter\ifx\csname r@fnum#1\endcsname\relax}%
\def\newr@f#1{\global\advance\r@fcount by1%
    \expandafter\xdef\csname r@fnum#1\endcsname{\number\r@fcount}}

\let\r@fis=\refis                       
\def\refis#1#2#3\par{\ifuncit@d{#1}
   \newr@f{#1}%
   \w@rnwrite{Reference #1=\number\r@fcount\space is not cited up to now.}\fi%
  \expandafter\gdef\csname r@ftext\csname r@fnum#1\endcsname\endcsname%
  {\writer@f#1>>#2#3\par}}

\def\ignoreuncited{
   \def\refis##1##2##3\par{\ifuncit@d{##1}%
     \else\expandafter\gdef\csname r@ftext\csname
r@fnum##1\endcsname\endcsname%

     {\writer@f##1>>##2##3\par}\fi}}

\def\r@ferr{\endreferences\errmessage{I was expecting to see
\noexpand\endreferences before now;  I have inserted it here.}}
\let\r@ferences=\references
\def\references{\r@ferences\def\endmode{\r@ferr\par\endgroup}}

\let\endr@ferences=\endreferences
\def\endreferences{\r@fcurr=0
  {\loop\ifnum\r@fcurr<\r@fcount
    \advance\r@fcurr by 1\relax\expandafter\r@fis\expandafter{\number\r@fcurr}%
    \csname r@ftext\number\r@fcurr\endcsname%
  \repeat}\gdef\r@ferr{}\endr@ferences}


\let\r@fend=\endpaper\gdef\endpaper{\ifr@ffile
\immediate\write16{Cross References written on []\jobname.REF.}\fi\r@fend}

\catcode`@=12

\citeall\refto          
\citeall\ref            %
\citeall\Ref            %



\def \bk {{\bf k}}
\def \bkp {{\bf k}'}

\hfill{ LA-UR 91-3232}

\hfill{RU 91-48}
\title A novel class of singlet superconductors.
\vskip .2truein
\author Alexander Balatsky$^{\dag}$

\affil Los Alamos National Laboratory,
Center for Materials Science, MS K765
Los Alamos, NM 87545

\author Elihu Abrahams

\affil Serin Physics Laboratory,
Rutgers University
P.O. Box 849, Piscataway, NJ 08854

\vskip 0.2 truein
\abstract{ A new class of singlet superconductors with a gap function
$\Delta(\bk, \omega_n)$ which is {\it odd} in both momentum and
Matsubara frequency is considered. Some of the physical properties
of this superconductivity are discussed and it is argued that: i) the
electron-phonon interaction can produce this kind of pairing, ii) in
many cases there is no gap in the quasiparticle spectrum, iii) these
superconductors will exhibit a Meissner effect.}

\vskip .4in
\noindent PACS Nos. 74.20-z; 74.65+n; 74.30 Ci.
\vskip .1 truein
\dateline

\endtopmatter
\body

Some recent models of high-T$_c$ superconductors with
unusual structure of the gap function $\Delta
({\bf k}, \omega_n$) have introduced general questions about the
possible symmetry types of the gap
for singlet superconductors. For example, Mila and
Abrahams\refto{MA} discussed a singlet superconductor with a
gap which is an odd function in ($k-k_F$). This form, as discussed by
Anderson,\refto{pwarg} annihilates the effect of strong short-range
repulsion.

A careful symmetry analysis
leads us to the conclusion that in addition to the standard BCS-like
singlet gap function, there is a
new, apparently unnoticed, class of singlet superconductors,
whose gap function  $\Delta
({\bf k}, \omega_n$) and
anomalous Green's function are {\it odd} in both Matsubara
frequency
$\omega_n$ and momentum ${\bf k}$.

Nearly two decades ago in a little-noticed article,
Berezinskii\refto{ber} considered the possibility
of unusual $S=1$ triplet pairing in $^3$He. He argued that it is
permissible, from the point of
view of symmetry of the superconducting gap, to have a phase in
which the gap function is a vector in
spin space for triplet case, ${\bf \Delta} (\omega_n, {\bf k}$), odd in
Matsubara
frequency and even in momentum ${\bf k}$. Although it is now
commonly believed that, in the observed phases, the
gap in superfluid $^3$He is even in frequency and odd in ${\bf k}$,
there is no
symmetry restriction which prohibits the phase proposed by
Berezinskii.

We shall adapt Berezinskii's approach\refto{ber} to the singlet case.
We introduce the anomalous Green's function in $d$-dimensions
$$
F({\bf k}, \omega_n)~=~{1\over 2}\sum_{\alpha,\beta}\int d{\bf r}~\int^
{\beta}_{-
\beta}~d\tau~e^{i\omega\tau}~e^{i{\bf k}\cdot{\bf r}}~\langle
T_\tau\,\psi_\alpha (\tau,{\bf r})~\psi_\beta (0,0)\rangle
g_{\beta\alpha} \eqno(1)
$$

\noindent
with the notations: $g_{\alpha\beta} = (i\sigma_y)_{\alpha\beta}$ is
a spin metric
tensor, $\tau$ is the Matsubara time, and $\beta = 1/T$.
Note that the anomalous Green's
function is explicitly written in a general spin-singlet form; the
function
$F({\bf k},\omega_n)$ is a true
scalar: $S^+ F({\bf k},\omega) \equiv 0$, where $S^+ =
\displaystyle\sum S^+_i$ is the total spin-raising operator. The
same discussion holds for the anomalous self energy $W({\bf k}, \omega_n)$
and the gap function $\Delta ({\bf k},\omega_n)$.\refto{zeven}

If one {\it assumes} that the spatial wave function for the singlet
Cooper pair is an even
function under ${\bf k} \to - {\bf k}$, the standard BCS expression
for the gap $\Delta
({\bf k},\tau) \propto \langle T_\tau \psi_{{\bf k}\uparrow} \psi_{-\bf
{k}\downarrow}\rangle$ is recovered. We do
not want to make any assumptions at this point, so the
anomalous Green function $F$ and anomalous self energy $W$
are taken in the form of the general singlet, Eq.\ (1).
Then the {\it only} constraint on the possible symmetry of $F$ and $W$
follows from the anticommutativity of the $\psi$ operators in $F$,
and we immediately get, for the singlet case:\refto{zeven}
$$
\eqalignno{F ({\bf k},\omega_n)~&=~F (-{\bf k}, -\omega_n) &(2a)\cr
\noalign{\vskip 0.5pc}
\Delta ({\bf k},\omega_n)~&=~\Delta (-{\bf k}, -\omega_n) &(2b)\cr}
$$

\noindent
There are two distinct ways to satisfy Eqs.\ (2) in terms
of definite symmetry
types of the gap:

(a) The standard Eliashberg-BCS singlet gap which is even both in
$\omega_n$ and ${\bf k}$: $\Delta ({\bf k},\omega_n) = \Delta (-{\bf
k},\omega_n) = \Delta ({\bf k}, -
\omega_n$). For this kind of pairing the equal time anomalous Green
function is nonzero,
leading to the usual off-diagonal long-range order, ODLRO. Then the
equal time Cooper pair
orbital wave function has to be symmetric in electron coordinates
since the spin wave function
is a singlet and antisymmetric.

(b) Singlet superconducting
pairing with a gap which is odd in both  ${\bf k}$ and $\omega_n$:
$$
\eqalignno{F ({\bf k}, \omega_n)~&=~- F (-{\bf k}, \omega_n)~=~- F
({\bf k}, -
\omega_n)&(3a)\cr
\noalign{\vskip 0.5pc}
\Delta ({\bf k}, \omega_n)~&=~- \Delta (-{\bf k}, \omega_n)~=~-
\Delta ({\bf k}, -
\omega_n).&(3b)\cr}
$$

\noindent
In this note, we shall consider this novel kind of singlet
superconductivity.
Eq.\ (3b) implies that the spin-singlet gap is described in terms of an
odd orbital function,
while, at the same time, the spin function is odd. There is no violation
of the Pauli principle
because the equal time gap function vanishes since the gap is odd in
$\omega_n$.\refto{BCS} The physical consequences of this behavior
of the gap are far-reaching. For example, such a system does not
exhibit conventional ODLRO which
requires a nonzero equal-time anomalous correlator.

Before discussing the physical propertes of such a superconductor,
we
consider the microscopic Eliashberg equations which lead to this kind
of gap
function. With standard Nambu-Eliashberg notation, the matrix Green
function has the form:
$$
\hat G ({\bf k}, \omega_n)~=~{i\omega_n Z_{{\bf k}} (\omega_n)
\tau_o + W ({\bf k}, \omega_n)
\tau_1\over \omega_n^2 Z^2_{\bf k} (\omega_n) + |W ({\bf k},
\omega_n)|^2 + \epsilon^2_{\bf k}} \eqno(4)
$$

\noindent
The one loop self energies in the normal
and superconducting channels are:
$$
W ({\bf k}, \omega_n)~=~-~T~\sum_{n',{\bf k'}}~V_{\bf
{kk'}} (\omega_n -
\omega_{n'})~{W (\bkp, \omega_{n'})\over \omega_{n'}^2 Z^2_{\bf
k'} (\omega_{n'}) +
\epsilon^2_{\bf k'} +
|W ({\bf k'}, \omega_{n'})|^2}  \eqno(5a)
$$
$$
[1-Z_{\bf k}(\omega_n)]i\omega_{n}~=~T~\sum_{n',{\bf k'}}~V_{\bf{
kk'}}
 (\omega_n -
\omega_{n'})~
{i\omega_{n'} Z_{\bf k'}
(\omega_{n'})\over \omega_{n'}^2
Z^2_{\bf {k'}} (\omega_{n'}) +
\epsilon^2_{\bf k'} +
|W ({\bf k'}, \omega_{n'})|^2}, \eqno(5b)
$$

\noindent
where $V_{{\bf kk'}} (\omega_n - \omega_{n'})$ is some effective
interaction. These equations
are written with the assumption that the same interaction enters into
both Eqs.\ (5ab); the effect of impurities is neglected. It
follows from Eqs.\ (3ab)
that only the odd components in ${\bf k}, {\bf k'}, \omega_n$, and
$\omega_{n'}$ of the potential $V_{\bf kk'} (\omega_n
- \omega_{n'})$
contribute in the momentum integral and frequency sums in
Eq.\ (5a).
As indicated earlier,\refto{zeven} we assume in this paper that $Z_{\bf k}
(\omega_{n})$
is an even function of
${\bf k}$ and $\omega_n$. Other possibilities will be discussed in a subsequent
paper.\refto{ab} Then only the even-in-$\bk$ and odd-in-$\omega_n$ components
of
$V_{{\bf kk'}}
(\omega_n
- \omega_{n'})$ enter the RHS of Eq.\ (5b). The
${\bf k}$-dependence of the normal self energy near the Fermi surface is
usually weak, so we shall neglect
it in Eq.\ (5). We see that there are no intrinsic
inconsistencies within
the Eliashberg formulation which forbid the odd gap solution of
Eq.\ (3b).

In what follows, we discuss how an interaction mediated by phonons
can lead to the odd gap:
$$
V_{{\bf kk'}} (\Omega_m)~=~\alpha^2\,D_{{\bf kk'}} (\Omega_m) =
\alpha^2{2\over
\pi}~\int d\omega~{A_{{\bf kk'}} (\omega) \omega\over \omega^2 +
\Omega_m^2}, \eqno(6)
$$
where $\Omega_m=\omega_n-\omega_{n'}$ is an even (bosonic)
Matsubara frequency.
\noindent
Then antisymmetrization in $D_{{\bf kk'}} (\omega_n -
\omega_{n'}$) over
$\omega_{n'}$
automatically implies antisymmetrization over $\omega_n$. In the
phonon case, there needs to be sufficient ${\bf k}$-dependence in
$D_{{\bf kk'}}
(\Omega)$ to be able to produce odd in ${\bf k}, {\bf k}'$
interactions. Phonons do not contribute to the (odd) pairing kernel
of Eq.\ (5a) if they are described in the Einstein approximation with
${\bf k}$-independent
spectral density $A(\omega)$.\refto{varma}

To illustrate, consider the weak coupling ($Z = 1$) limit of the
Eliashberg Eqs.\ (5).
Although interaction with phonons does produce a $Z$-factor
renormalization, we neglect it for this discussion. Assume
that pairing is mediated by acoustic phonons with :
$$
V_{{\bf kk'}} (\Omega)~=~ \alpha^2  \ {c^2(\bk - \bkp)^2 \over
c^2(\bk - \bkp)^2
 + \Omega^2}.
\eqno(7)
$$
For $\bk \sim \bkp \sim k_F$ the frequency in the phonon
propagator is
usually small
in comparison with the term containing the momenta: $|\Omega| << c|
\bk -
\bkp|$. This allows us to expand $V_{{\bf kk'}} (\Omega)$ in Eq.\ (7).
Keeping
in mind that
only the odd in $\bk, \bkp, \omega_n$ and $\omega_{n'}$
components
contribute to the gap Eq.\ (5a), we get:
$$
V_{{\bf kk'},odd}~=~4\alpha^2{\bk\cdot\bkp
\omega_n\omega_{n'} \over c^2(\bk+\bkp)^2(\bk-\bkp)^2}~+~{\rm
O}([{\omega_c
\over ck_F}]^2),
$$
where $\omega_c$ is the maximum phonon frequency. The
linearized gap equation is then
$$
\Delta(\bk, \omega_n) = (4\alpha^2 T/c^2)\sum_{n', \bkp} {\bk\cdot
\bkp
\  \omega_n  \
\ \omega_{n'} \over (\bk^2 +  \bkp^2)^2 - 4(\bk\cdot \bkp )^2} \cdot
\
{\Delta(\bkp,
 \omega_{n'}) \over \omega_{n'}^2 + \epsilon_{\bkp}^2}. \eqno(8)
$$
 From Eq.\ (8), it follows that the gap has to be linear in frequency
up to
the cutoff $\omega_c$. We shall use the {\it Ansatz} :
$$
\Delta ({\bf k}, \omega_n)~=~{i\omega_n \over \omega_c} ~  {{\bf k}
\over k_F}\cdot{\bf d}
({\bf k},\omega_n) \eqno(9)
$$
with  ${\bf d}
({\bf k},\omega_n)  = {\bf d} \  \Theta(\omega_c-|\omega_n|)$,
where $\Theta(x)$ is
a step function.
Combining
Eqs.\ (8) and (9) when $T<\omega_c$,  we find that the gap equation
exhibits nontrivial solutions {\it above} a critical temperature $T_{c_{-}}$,
where
$$
1={\alpha^2 \over \alpha_c^2}(1+{3\over 2}\pi^2{T_{c_-}\over \omega_c}).
\eqno(10)
$$
Here $N_0\alpha_c^2  =  a(c k_F /\omega_c)^2$, where $a$ is a positive
constant of order unity.


The thermodynamics of this phase is different from the one for  BCS
superconductors: For intermediate couplings $\a^2  < \a^2_c$, the gap
equation leads to a nontrivial solution in the temperature range
$T_{c_+} > T > T_{c_-}$, where $T_{c_+}$, of order $\omega_c$, is the
temperature at which the smallest value of $\omega_{n'}$ in the sum of
Eq.\ (8) exceeds the cutoff, rendering the RHS zero.  In the region just above
$T_{c_-}$, the system is described by a Ginzburg-Landau
(GL) theory with order parameter $|{\bf d}| \propto (T - T_{c_-})^{1/2}$.
At larger values of the coupling, $\a^2  > \a^2_c$,  the
lower critical temperature $T_{c_-}$ goes to zero and the lower GL
region vanishes. Berezinskii\refto{ber} found
analogous results in his treatment of the odd-frequency gap for
triplet pairing.  Detailed analysis of the thermodynamics and GL
theory of this phase will be given elsewhere.\refto{ab}

A special case of the odd gap will occur if the form of the interaction
admits a solution of the form $\Delta(\bk,\omega_n) = \bk\cdot{\bf d}~{\rm
sgn}(n)$.
In this case, the $T_c$ equation from Eq.\ (5a) is precisely that of a
$p$-wave BCS superconductor and the condensed phase occurs for $T_c>T>0$.
We shall not discuss this possibility further.

We conclude that the only
criterion for a physical
system to choose between odd and even gaps is the overall minimum
of the free energy. From our discussion, it follows that the standard
BCS $s$-wave superconductivity will have lower energy, at least for
a weak electron-phonon interaction. However,
if one takes a strong short-range repulsion (as in the Hubbard model)
into consideration,\refto{pwarg} the ``no-double-occupancy"
constraint
$\sum_{k, \omega} \Delta(\bk, \omega) = 0$ must be obeyed in the
superconducting state. This is automatically
satisfied for
the odd gap and in this case, odd pairing may be favored over the
conventional BCS state whose energy will be raised by the repulsion.

Let us consider some of the physical properties of an odd gap
superconductor. An important consequence of a gap which is odd
under
$\tau \to -\tau$ and under
 ${\bf r} \to - {\bf r}$ is that one has broken time
reversal and parity. This leads to existence of the orbital Goldstone
vector ${\bf d}({\bf k}, \omega_n)$
which is analogous to the orbital momentum vector in the triplet
superconductors.  Below we will assume that ${\bf d}$ is a real
vector; however there are other possibilities.\refto{Aphase}

With our {\it Ansatz}, the quasiparticle spectrum for
such a superconductor
is gapless. Indeed, if we assume the gap function has the form in Eq.\ (9) with
a real ${\bf d} ({\bf k}, \omega_n)$
which is smooth and even in $\omega_n$ and ${\bf k}$,
then we find from the poles of the Green's
function, Eq.\ (4) (in weak coupling, $Z \simeq 1$), that
$$
\omega_k~\simeq~{\epsilon_k\over \sqrt {1+ (\bk \cdot {\bf
d})^2/(k_F\omega_c)^2}}.
\eqno(11)
$$
\noindent
Thus quasiparticle excitations in such a superconductor are gapless;
the only effect of
superconducting correlations is an effective mass renormalization,
$m_{\bk}^* = m\sqrt{1+ (\bk \cdot {\bf d})^2/(k_F\omega_c)^2}$. From
this point of view this
superconductor is essentially
a normal metal with nonlocal superconducting correlations. Note that
the gap vector ${\bf d({\bf k})}$ and the mass renormalization
vanish
when $\bk \perp {\bf d}$.
The gain in free energy in the superconducting state is given by the
standard BCS expression,\refto{AB}
$$
{\cal F}_s - {\cal F}_n  ~= ~  - T\sum_{\omega_n,\bk}  \int_0^1 d
\lambda
{|\Delta({\bf k},\omega_n)|^2 \over \omega_n^2 + \lambda^2
|\Delta({\bf k},\omega_n)|^2 + \epsilon_{{\bf k}}^2} \simeq   - {1
\over 2}
  N_0  \  {\bf d}^2,
\eqno(12)
$$
where the gap is assumed to have the form of Eq.\ (9) with
${\bf d} ( {\bf k}, \omega_n)$ independent of $\omega_n$ and
where $N_0$ is the density of
states at the Fermi surface. This formula also follows from the
observation that the
effect of such pairing on the low energy states is an increase of the
density of states $N^*
 =  N_0  m^*/m$. This results in an
energy change
 $ \delta E =  \omega^2_c (N^* - N_0)$  which is equal to the r.h.s. of
Eq.\ (12).

There is no static order
parameter since $F({\bf r}_1,{\bf r}_2; t_1,t_1)=0$. Nevertheless, the
global
electromagnetic $U(1)$
group is broken because even for nonequal times $t_1,t_2$ and space
points $ {\bf r}_1 , {\bf r}_2$, the existence of the anomalous
correlator implies $\langle \psi_{\alpha}(t_1, {\bf r}_1) ~
\psi_{\beta}(t_2, {\bf r}_2) \rangle \rightarrow   e^{ i 2 \phi }\langle
\psi_{\alpha}(t_1, {\bf r}_1) ~ \psi_{\beta}(t_2, {\bf r}_2) \rangle $
under this transformation. This suggests that the electromagnetic
response of these superconductors will be the same as for BCS
superconductors; they will exhibit a Meissner effect. In order
to calculate the kernel in  linear response, we shall use
standard expressions from the BCS theory,\refto{AGD} and take into
account the frequency and momentum dependence of the gap.
Because the gap function is a scalar in our case, the correction to the gap
function $ \Delta({\bf k},\omega_n)$ which is linear in the gauge
potential ${\bf
A}$, is proportional to div${\bf A}$. In the gauge div${\bf A} = 0$,
we can use the linear response theory with the unperturbed gap function
given by Eq.\ (9). This can be checked within linear response theory
directly with the use of the Peierls substitution $ \bk
\rightarrow \bk +
 2e{\bf A}$.
 The
kernel for the static response has the form:
$$
Q( {\bf k})~ = 1 + {3\over4} T \sum_{\omega}~\int_0^{\pi} d\theta
sin^3 \theta \int_{-\infty}^{\infty}d \xi
{( i \omega + \xi_-)( i \omega + \xi_+) + \Delta_- \Delta^*_+ \over {(
\omega^2 + \xi^2_- + |\Delta_-|^2) (\omega^2 + \xi^2_+ + |\Delta_+|^2)}},
\eqno(13)
$$
where $\xi_{\pm} = \xi \pm {1 \over 2 }  {\bf k}\cdot{\bf v}$ and
analogous notations for the gap. From Eq.\ (13), we can find the
asymptotic kernel for small momenta, assuming that the gap
$\Delta({\bf
k}, \omega_n)$ is essentially linear in momentum and frequency as
in Eq.\ (9):
$$
Q(k \rightarrow 0) \simeq  \  {\pi \over 2 }\ln{\omega_c \over T} \
{
({\bf d} /\omega_c)^2 \over ( 1 +  ({\bf d} / \omega_c)^2 )^{3/2}}
\eqno(14)
$$
The fact that the
kernel is logarithmically divergent means that this particular type of
superconductor is of the Pippard type at low enough temperatures
(the temperature has to be very small because of the weak
logarithmic
divergence). In the vicinity of the critical temperature, however, the
temperature dependence of the penetration depth is that of the gap
squared:
$$ \lambda^2 = {2 m \over N e^2 \pi} {( 1 +  ({\bf d} / \omega_c)^2
)^{3/2}
\over ({\bf d} /\omega_c)^2 }  \sim {1 \over {\bf d}^2}. \eqno(15)
$$
This makes this superconductor of the London type in the vicinity of
$T_{c_-}$.
If we assume that the gap as a function of frequency
has larger power than unity, we can get a penetration depth
which is finite in the whole range of temperature.

In conclusion we found a new class of singlet superconductors with a
gap which is {\it odd} in both momentum and frequency and we showed that
there is no symmetry restriction which prohibits this kind of gap
function. The physical properties of these superconductors are rather
unusual.  Parity and time  reversal symmetries are broken; this leads
to Goldstone modes and makes these singlet superconductors
analogous to superfluid $^3$He. There is no gap in the quasiparticle
spectrum, and the equal-time anomalous (pair) correlator vanishes.
Hence, there is no ODLRO in the usual sense but we find that there is
a Meissner effect. Static impurity scattering will be pair-breaking,
as is usual for anisotropic superconductors. At moderate coupling,
the normal phase reenters below $T_{c_-}$. The coherent state appears
to be a result of pairing among the thermally excited quasiparticles
which are present at non-zero temperature. All these nontrivial
properties deserve further
investigation.

\vskip .2in
We acknowledge useful correspondence with P.B. Allen.
This work
was supported in part by NSF grant DMR 89-06958 (E.A.) and by a
J.R.\ Oppenheimer fellowship and the Department of Energy (A.B.),
and by the Advanced Studies Program of the Center for Materials
Science at Los Alamos National Laboratory.

\vskip 2pc

\noindent
\endpage

\references

\noindent
$^{\dag}$ On leave from Landau Institute for Theoretical Physics,
Moscow, USSR.

\refis{MA} F. Mila and E. Abrahams, Phys. Rev. Lett., October 21,
1991. See also S. Nakajima, Prog.\ Theor.\ Phys.\ {\bf 32}, 871 (1964);
M.H. Cohen, Phys.\ Rev.\ Lett.\ {\bf 12}, 664 (1964).

\refis{pwarg} P.W. Anderson, Workshop on Fermiology of High-$T_c$
Superconductivity, Argonne, March 1991 (J. Phys.
Chem. Solids).

\refis{ber} V. L. Berezinskii, JETP Lett.\ {\bf 20}, 287 (1974); [ZheTP
Pisma {\bf 20}, 628 (1974)].

\refis{BCS} Note also that the odd gap {\it cannot} be obtained within
the weak coupling BCS approximation
which assumes no frequency dependence of the gap $\Delta ({\bf k},
\omega_n)$.

\refis{AGD} A. Abrikosov, L.P. Gor'kov and I. E. Dzyaloshinskii,
{\sl Methods of Quantum Field Theory in Statistical Physics}, Dover,
New
York (1975).

\refis{AB} A.V. Balatsky and V. P. Mineev, Sov.\ Phys.\ JETP {\bf 62},
1195 (1985).

\refis{varma} A. J. Millis, S. Sachdev and C. M. Varma, Phys.\ Rev.\
B37,
4975 (1988).

\refis{Aphase} In principle ${\bf d}(\bk, \omega)$ can be complex
vector, say ${\bf d} \propto {\bf e_x} + i{\bf e_y}$ as
in $^3$He-$A$. In
this case
the gap will have nodes at $\bk \perp {\bf d}$ at two
points on the Fermi surface. The broken time reversal and
parity symmetries lead
to the existence of an intrinsic orbital momentum ${\bf
L}_0\parallel{\bf d}$ of
order of
$({\bf d}
 / E_F)^2 $. The Goldstone modes associated with the
motion of the
${\bf d}$ vector are additional low-lying modes which also
distinguish this
superconductor from the usual BCS singlet superconductor.

\refis{zeven} Recall, $W$ and $\Delta$ are related by
$\Delta=W/Z$, where $Z$ is the self energy in the normal channel, which
in this paper we
shall take to be even in $\bk, \omega_n$.

\refis{ab} E. Abrahams and A. V. Balatsky, unpublished.

\endreferences


\end